\documentclass[runningheads]{llncs}
\usepackage[T1]{fontenc}
\usepackage{graphicx}
\usepackage{tikz}        
\usepackage{pgfplots}   
\usepackage{graphicx}
\begin{document}
\title{ABACUS: A FinOps Service for Cloud Cost Optimization}
%
%
\author{Saurabh Deochake\orcidID{0000-0002-3757-6463}}
\authorrunning{Saurabh Deochake}
%
\institute{SentinelOne Inc.\\ 444 Castro St, Mountain View, CA 94041 USA\\
\email{saurabh.deochake@sentinelone.com}}
\maketitle              
\begin{abstract}
In recent years, as more enterprises have moved their infrastructure to the cloud, significant challenges have emerged in achieving holistic cloud spend visibility and cost optimization. FinOps practices provide a way for enterprises to achieve these business goals by optimizing cloud costs and bringing accountability to cloud spend. This paper presents ABACUS - Automated Budget Analysis and Cloud Usage Surveillance, a FinOps solution for optimizing cloud costs by setting budgets, enforcing those budgets through blocking new deployments, and alerting appropriate teams if spending breaches a budget threshold. ABACUS also leverages best practices like Infrastructure-as-Code to alert engineering teams of the expected cost of deployment before resources are deployed in the cloud. Finally, future research directions are proposed to advance the state of the art in this important field.

\end{abstract}
\section{Introduction}\label{sec:introduction}
In today's fast-changing digital landscape, cloud computing has become an essential tool for businesses looking to promote innovation, improve agility, and achieve scalability. Organizations across industries are progressively shifting to the cloud, drawn by the promise of increased performance and lower infrastructure expenses. Therefore, with the increasing growth of public cloud infrastructure hosting, conventional on-premise data center-based businesses are shifting their workloads to the public cloud. Compared to traditional data center infrastructure, public clouds offer greater elasticity, efficiency, and scalability for Infrastructure as a Service (IaaS). However, the spike in cloud use frequently results in unexpected obstacles, particularly in monitoring and controlling cloud expenses. As cloud services become more prevalent, the complexities of invoicing and resource allocation also become commonplace. This makes it increasingly difficult for businesses to maintain a tight grip on their financial management. 

Cost optimization, therefore, has become an increasingly important concern for organizations of all sizes as cloud computing adoption continues to grow year by year. According to a Flexera report, in recent years, the main cloud initiative for businesses has been cost optimization \cite{flexera_2024}. Furthermore, Gartner predicts that "organizations that do not have a cost optimization program in place will overspend by up to 70\%" through 2024 \cite{gartner_research_2021}. The impacts of failing to manage cloud costs are obvious: wasted resources, increased expenses, and lower overall profitability. Therefore, understanding and implementing effective cloud cost optimization strategies is critical for businesses to remain competitive in the modern cloud computing landscape \cite{deochake_2023}.

To manage and optimize these operating expenses (OpEx) costs, the concept of Financial Operations (FinOps) emerges as a game changer. FinOps is a collaborative method that bridges the gap between finance, engineering, and operations teams, allowing organizations to efficiently control cloud expenditures. FinOps enables teams to make informed decisions about cloud resource utilization by cultivating an accountability and transparency culture, ensuring that expenditure is in sync with business objectives.

This paper proposes ABACUS - Automated Budget Analysis and Cloud Usage Surveillance, an automated FinOps solution developed specifically to address the challenge of cloud cost optimization. ABACUS leverages automation and data analytics to give enterprises access to cloud cost attribution, budgeting, alerts, and infrastructure spend. Furthermore, the solution helps teams discover unproductive cloud resources and recommend smart cost-cutting methods, such as temporarily blocking new cloud workloads when budgets are exceeded. ABACUS easily connects with financial planning data, including OpEx costs and budgets. By integrating with existing cloud providers and utilizing cost-attribution techniques such as resource tagging, it also translates and assigns budget data to cloud resources belonging to a specific cost center or department.  

The structure of this paper is as follows. Section \ref{sec:cloud_cost} delves into the critical necessity for cloud cost optimization. Section \ref{sec:finops} explores in depth the guiding principles of FinOps. Furthermore, Section \ref{sec:abacus} introduces ABACUS as a system that applies the key tenets of FinOps to the challenge of cost minimization. Section \ref{sec:future_work} examines the future work recommended to enhance the solution with novel features. Finally, Section \ref{sec:conclusion} concludes the paper.

\section{The Need for Cloud Cost Optimization} \label{sec:cloud_cost}
As organizations increasingly use cloud computing, related operating expenses (OpEx) have skyrocketed. As per a recent report, global cloud computing market was valued at \$626.4 billion in 2023 and is expected to increase to \$1,266.4 billion by 2028, with a compound yearly growth rate (CAGR) of 15.1\% during this period \cite{markets_and_markets_2023}. This rapid expansion emphasizes the necessity for organizations to manage and optimize their cloud expenditures efficiently. 

The dynamic and scalable nature of cloud resources makes cost management one of the most difficult tasks. Unlike traditional on-premises infrastructure, cloud services can be rapidly scaled up or down in response to demand, complicating cost forecast and control \cite{bib2} \cite{bib3}. For example, in response to sudden increase in user traffic, the automatic resource scaling can boost performance, but it may also lead to unexpected cost spikes if not closely monitored. Furthermore, the complexity of cloud billing, which often includes both fixed and variable costs, makes it difficult for businesses to accurately estimate and budget for cloud expenses \cite{deochake_2023}. This lack of transparency can result in unclear spending and misalignment with business goals.

A lack of visibility into cloud spending can result in inefficient resource allocation and significant loss, with data from a recent Acceldata report indicating that businesses can lose up to 32\% of their cloud budget on idle or underutilized resources \cite{shaikh_2024}. This issue frequently occurs as organizations seek to identify and eliminate inefficiencies, especially in environments where several teams provision and maintain cloud services separately. Such a decentralized strategy can result in information silos, making it difficult to acquire a full knowledge of cloud expenditures and resource utilization across the firm. Without this information, engineers and IT management may struggle to successfully optimize resources, match spending with business goals, or even determine where the most substantial cost-saving potential exist. As a result of this, the potential for waste escalates, and the business risks overspending on cloud services that do not deliver proportional value.

Furthermore, the rapid expansion of cloud services usually results in shadow platforms where different departments and teams adopt cloud solutions without the knowledge or approval of the central platform and finance teams. This lack of coordination might result in a fragmented approach to cloud cost management, with expenses that are not managed or optimized efficiently. Failure to adequately manage cloud expenses can result in major consequences, ranging from financial strains and budget overruns to missed chances for research and development. According to a recent white paper by Civo, 37\% of organizations faced unanticipated cloud charges, highlighting the crucial need for enhanced cost management strategies \cite{civo_2024}. When departments work in silos, the lack of visibility and control over cloud spending can quickly lead to a situation where expenses are unexpected and resources are allocated inefficiently. Therefore, this not only complicates budget planning but also hampers the ability to support strategic business initiatives, as funds that could drive innovation are instead absorbed by unmonitored and unnecessary expenses.

Therefore, as businesses rely more on cloud services to power their digital transformation, efficient cloud cost optimization measures become critical. Implementing strong cost management techniques and leveraging modern tools can provide organizations with critical control over their cloud spending. This includes establishing governance frameworks to ensure accountability, utilizing tagging and resource management strategies to improve visibility, and implementing automated tools to discover and resolve resource inefficiencies in real time. To summarize, cloud cost optimization is more than simply a financial imperative; it is a strategic requirement for enterprises seeking to properly embrace cloud computing. The capacity to manage costs while maximizing the value of cloud investments is important for preserving a competitive advantage in an increasingly digital world. In the following sections, we will look at the ideas of FinOps and introduce ABACUS, a comprehensive solution that may assist enterprises negotiate the intricacies of cloud cost optimization and achieve long-term financial sustainability.

\section{Understanding FinOps}\label{sec:finops}
\subsection{What is FinOps}
FinOps, or Financial Operations, is an emerging discipline at the intersection of finance and engineering that tackles the complicated task of controlling and optimizing operating expenses (OpEx) in today's quickly changing IT ecosystem. As organizations rely more on cloud services, the need for a structured approach to cloud financial management has become critical.

At its core, FinOps is about bringing financial accountability to the variable spend and cost model of cloud computing. It is a collaborative effort that brings together technology, finance, and business teams to make informed decisions about cloud usage and expenditure. The FinOps Foundation which is a program of the Linux Foundation defines it as "an evolving cloud financial management discipline and cultural practice that enables organizations to get maximum business value by helping engineering, finance, technology and business teams to collaborate on data-driven spending decisions" \cite{finops_foundation_2023}. Therefore, as per the FinOps Foundation, the primary goals of FinOps include the following.

\begin{itemize}
    \item Improving visibility into cloud costs across multi-cloud environments
    \item Optimizing resource utilization and eliminating idle resource waste
    \item Aligning cloud spending with business financial objectives
    \item Creating a culture of financial accountability among all cloud stakeholders
\end{itemize}

FinOps involves setting up the right processes, tools, and organizational structures to help teams make informed decisions about cloud resource allocation and usage. This includes managing costs, budgeting effectively, detecting anomalies, and employing predictive analytics to keep spending in check. However, FinOps is not solely focused on cutting costs. Instead, it emphasizes making strategic trade-offs between speed, cost, and quality. For instance, there are times when investing more in cloud resources can lead to significant business benefits, such as faster time-to-market or enhanced performance. Conversely, there are situations where cost optimization and efficiency become critical to maintaining financial health.

Adopting FinOps principles means aiming for a balanced approach that aligns innovation with operational efficiency and cost management. It involves setting up robust governance frameworks and leveraging tools that provide visibility into cloud spending, enabling teams to make data-driven decisions. This strategic approach not only helps in controlling costs but also ensures that cloud resources are utilized in ways that drive business value. As a result, organizations can enhance their operational effectiveness, adapt to market changes more rapidly, and secure a competitive advantage in an increasingly cloud-centric landscape \cite{storment2023cloud}.

\subsection{Core Principles of FinOps}
FinOps is built on several core principles that guide organizations in effectively managing their cloud financial operations. These principles foster collaboration, accountability, and strategic decision-making across teams involved in cloud spending \cite{finops_foundation_2023}. The core principles of FinOps can be thought of as following.

\begin{itemize}
    \item \textbf{Collaboration Across Teams}: FinOps emphasizes the importance of collaboration between finance, engineering, and operations teams. By breaking down silos, all stakeholders can contribute to cloud financial management, ensuring that everyone understands the financial implications of their decisions. This collaboration allows teams to align on cloud spending priorities, identify cost optimization opportunities, and make informed trade-offs between cost, speed, and quality.
    \item \textbf{Centralized FinOps Team}: While collaboration is key, having a centralized FinOps team helps to drive the overall strategy and governance of cloud financial management. This team acts as a resource for best practices, tools, and processes, ensuring consistency across the organization. The FinOps team also plays a crucial role in educating and training other teams on cloud financial management, fostering a shared understanding of FinOps principles and methodologies.
    \item \textbf{Ownership and Accountability}: FinOps promotes a culture where all teams take ownership of their cloud usage and associated costs. This accountability drives teams to be more mindful of their resource consumption and encourages proactive management of cloud expenditures. By fostering a sense of ownership, organizations can reduce waste, optimize resource utilization, and instill a cost-conscious mindset among cloud consumers.
    \item \textbf{Leveraging the Cloud's Variable Cost Model}: One of the unique aspects of cloud computing is its variable cost model. FinOps encourages organizations to take advantage of this model by optimizing resource usage and exploring cost-saving opportunities, such as reserved instances or spot pricing. By embracing the flexibility of the cloud, organizations can scale resources up or down based on demand, avoiding the need for costly over-provisioning or under-provisioning of infrastructure.
    \item \textbf{Accessible and Timely Reporting}: For FinOps to be effective, financial data must be readily accessible and presented in a timely manner. This transparency allows teams to make informed decisions based on current spending patterns and budget constraints, enabling quick adjustments as needed. Accurate and up-to-date reporting also helps in identifying cost anomalies, tracking budget adherence, and communicating cloud financial performance to stakeholders.
\end{itemize}

This FinOps approach not only helps in controlling cloud costs but also enables teams to leverage cloud resources more strategically, ultimately driving better business outcomes.

\subsection{The FinOps Lifecycle}
The FinOps lifecycle is a comprehensive framework that helps organizations navigate the complexities of cloud financial management. Roughly, it consists of three iterative phases: Inform, Optimize, and Operate, each with its own set of objectives and activities. 

\begin{itemize}
    \item \textbf{Inform}: In the Inform phase, the focus is on establishing visibility and accountability by identifying data sources for cloud cost and usage metrics, accurately allocating expenses based on tags or business rules, and developing budgeting and forecasting capabilities. This phase empowers teams with accurate and timely data, enabling them to make informed decisions and align cloud spending with business objectives. 
    \item \textbf{Optimize}: The Optimize phase is dedicated to identifying and implementing opportunities to improve cloud efficiency and cost-effectiveness. Teams work collaboratively to rightsize underutilized resources, leverage cloud provider optimization options, and select the best optimization opportunities based on organizational goals. By optimizing resource utilization and exploring cost-saving opportunities, organizations can achieve more value from their cloud investments. 
    \item \textbf{Operate}: The Operate phase focuses on establishing organizational changes and a culture of cost accountability. Key activities include defining cloud governance policies, empowering individuals through training and automation, and continuously evaluating the alignment between business objectives and cloud spend. This phase ensures that cloud spending remains under control and adaptable to changing needs, fostering a culture of financial accountability across teams. 
\end{itemize}

ABACUS follows the similar principles of FinOps lifecycle in its operations by informing, optimizing, and operating the cost optimization operations in a multi-cloud environment.

\subsection{The Benefits and Challenges for FinOps}
As discussed in the sections above, FinOps provides a structured approach to managing cloud costs, enabling organizations to optimize their investments while navigating the complexities of cloud financial operations. However, while the benefits of implementing FinOps are significant, organizations also face various challenges that must be addressed to ensure successful adoption.

\subsubsection{Benefits of FinOps}
Implementing FinOps offers numerous benefits that enhance an organization’s ability to manage cloud financial operations effectively.
\begin{itemize}
    \item \textbf{Cost Savings}: One of the primary advantages is cost savings, as organizations can significantly reduce expenses through practices such as resource rightsizing, demand management, and leveraging various pricing models offered by cloud providers. This proactive approach eliminates waste from idle resources, allowing companies to reinvest those savings into other critical areas of the business. 
    \item \textbf{Enhanced Financial Performance}: Additionally, enhanced financial performance is achieved through improved visibility into cloud costs, enabling better budgeting, forecasting, and strategic decision-making.
    \item \textbf{Transparency}: The emphasis on transparency fosters a culture of accountability, where all stakeholders understand their cloud spending and its implications, leading to more informed resource allocation and project prioritization. 
    \item \textbf{Cross-functional Collaboration}: Furthermore, FinOps encourages cross-functional collaboration, breaking down silos between finance, engineering, and operations teams, which ultimately drives innovation and aligns cloud investments with overall business objectives.
\end{itemize}

\subsubsection{Challenges of FinOps}
Despite its many benefits, the implementation of FinOps also presents several challenges that organizations must address. This section briefly discusses some of major challenges that FinOps presents.

\begin{itemize}
    \item \textbf{Resistance to Change}: The first significant hurdle is resistance to change, particularly when transitioning to a collaborative approach that requires involvement from multiple departments. Teams may be accustomed to operating independently, making it difficult to establish a unified strategy for cloud financial management.
    \item \textbf{Complexity of Managing Costs}: Additionally, the complexity of managing costs in multi-cloud environments can complicate effective implementation, as organizations must navigate different pricing structures, billing models, and usage patterns across various platforms.
    \item \textbf{Ongoing Collaboration}: Another challenge is the need for ongoing collaboration between finance, engineering, and operations teams, which can be difficult to maintain over time. Ensuring that all stakeholders are engaged and informed throughout the FinOps process is crucial for success.
\end{itemize}

To overcome these challenges, organizations are encouraged to adopt a phased approach, gradually implementing FinOps practices while fostering a culture of financial accountability and collaboration. Balancing the benefits with these inherent challenges is essential for maximizing the value of cloud investments through FinOps. ABACUS aids in reducing the frictions mentioned above and streamlines the cloud cost optimization based on recommended FinOps principles.

\section{ABACUS}\label{sec:abacus}
This section showcases ABACUS, Automated Budget Analysis and Cloud Usage Surveillance, a service that follows FinOps principles and automates the analysis of budget for the cloud infrastructure and adjusts the cloud costs accordingly.

\subsection{Prerequisites}
Since ABACUS service depends on accurate budget and usage analysis of the cloud resources, it is imperative that certain prerequisites are established to ensure its effectiveness in managing cloud financial operations and costs.

\begin{itemize}
    \item \textbf{Resource Labeling}: First and foremost, resource labeling or resource tagging in certain cloud platforms is essential. Each cloud resource must be consistently labeled in form of key-value pairs that clearly identity their purpose, ownership, and environment at minimum. The practice of tagging or labeling the resources not only facilitates better organization but also ensures precise cost tracking and analysis. When resources are labeled, it becomes easier to attribute costs to specific teams or cost departments, allowing for more accurate budgeting and resource allocation. For example, \texttt{environment:prod} may accurately attribute costs to running the infrastructure and services. Whereas, \texttt{environment:dev} may accurately attribute the costs to research and development (R\&D). On shared platform services, however, it becomes trickier to attribute costs to only those parts of shared services that a team or cost department has used. To achieve granular cost attribution, specific resource labeling practices like labeling the end to end pipelines in such shared platforms should be practised. Resource labeling strategies are out of scope for this paper.
    \item \textbf{Cloud Accounts per Service}: Next, it is essential to establish GCP project, AWS accounts, or Azure subscriptions for each service or team. The accounts per service may be a better separation than account per team for organizations with changing priorities and frequent reorganizations. By creating dedicated accounts, organizations can ensure that each service or team operates within its own financial boundaries, making it easier to monitor spending and enforce accountability.
    \item \textbf{Allocation of Budgets}: The final prerequisite requires that the finance team must allocate budgets for each service or team. This involves setting clear financial parameters for cloud spending, ensuring that every team has a defined budget to work within. By establishing the budgets upfront, organizations can create a culture of financial responsibility and accountability. This, furthermore, encourages teams to optimize their resource usage and make informed decisions about their cloud investments.
\end{itemize}

\subsection{Budget Analysis Algorithm}
This section showcases how ABACUS systematically analyzes and sets budgets for cloud resources for a service or a team. This process is important for maintaining financial accountability and ensuring that the cloud spend aligns with organizational financial objectives.

\subsubsection{Analyzing Budgets}
The initial step in the budget analysis process involves gathering data on historical cloud spending (\texttt{HC}) and projecting future costs based on anticipated growth (\texttt{G}). To effectively manage budgets, ABACUS employs the following formula to calculate Cloud Resource Budget (CRB), which includes a variability factor (\texttt{V}) to account for potential fluctuations in spending:

\begin{equation}
CRB = \min \left( \sum_{i=1}^{n} HC_i \times (1 + G) \times (1 - C) \times (1 + V), AB \right)
\end{equation}

where,
\begin{itemize}
\item \textit{\texttt{CRB}}: Cloud Resource Budget
\item \textit{\texttt{HC\textsubscript{i}}}: Historical Cloud Spend for each instance or service, indexed by \textit{\texttt{i}} from 1 to \textit{\texttt{n}}
\item \textit{\texttt{G}}: Projected Growth Factor (expressed as a decimal, e.g., 20\% as 0.20)
\item \textit{\texttt{C}}: Cost Control Factor (expressed as a decimal, e.g., 10\% as 0.10)
\item \textit{\texttt{V}}: Variability Factor, which accounts for fluctuations in spending or growth projections (expressed as a decimal, e.g., 5\% as 0.05)
\item \textit{\texttt{AB}}: Available Budget
\item \textit{\texttt{min}}: The minimum function, ensuring that the calculated budget does not exceed the available budget
\end{itemize}

The variability factor \texttt{V} is calculated based on the historical spending data using the following steps:

\begin{enumerate}
    \item \textbf{Calculate the mean (\( \bar{x} \))} of the historical spending data:
    \[
    \bar{x} = \frac{1}{n} \sum_{i=1}^{n} x_i
    \]
    Where:
    \begin{itemize}
        \item \( n \) is the number of data points.
        \item \( x_i \) are the individual spending values for services or teams.
    \end{itemize}

    \item \textbf{Calculate the sample variance (\( s^2 \))} of the historical spending data:
    \[
    s^2 = \frac{1}{n - 1} \sum_{i=1}^{n} (x_i - \bar{x})^2
    \]

    \item \textbf{Calculate the standard deviation (\( s \))}:
    \[
    s = \sqrt{s^2}
    \]

    \item \textbf{Calculate the Coefficient of Variation (\( V \))} by dividing the standard deviation by the mean:
    \[
    V = \frac{s}{\bar{x}}
    \]
    
    The resulting \( V \) represents the relative variability of the data, expressed as a percentage of the mean.
\end{enumerate}

The formula serves several key purposes as mentioned below.
\begin{itemize}
\item \textbf{Dynamic Budgeting}: By incorporating historical spending, projected growth, and the variability factor, the formula allows organizations to create a budget that adapts to changing financial goals and needs while accounting for potential fluctuations.
\item \textbf{Cost Savings Measures}: The inclusion of the Cost Control Factor \texttt{C} enables organizations to apply strategic measures to manage and reduce costs, ensuring that the budgets reflect realistic spending limits.
\item \textbf{Financial Accountability}: The use of Available Budget \texttt{AB} as a minimization constraint ensures that the spending does not exceed the organizational limits, enforcing financial accountability among teams and cost departments.
\item \textbf{Data-driven Decisions}: The formula leverages historical spend data and variability calculations to inform future budgeting decisions, facilitating a data-driven approach to financial operations.
\end{itemize}
To showcase the application of this formula, let's consider the following hypothetical values for a team or service in an organization.
\begin{itemize}
\item Historical Cloud Spend \texttt{HC}: \$100,000
\item Projected Growth Factor \texttt{G}: 20\% (mentioned as 0.2)
\item Cost Control Factor \texttt{C}: 10\% (mentioned as 0.1)
\item Variability Factor \texttt{V}: 5\% (mentioned as 0.05)
\item Available Budget \texttt{AB}: \$120,000
\end{itemize}

Using the formula, we can then calculate the Cloud Resource Budget \texttt{CRB} for that team or service as follows:

\begin{enumerate}
    \item Calculate Adjusted Spend, \texttt{AS}: 
    \[
        \texttt{AS} = \newline
        \sum_{i=1}^{n} HC_i \times (1 + G) \times (1 - C) \times (1 + V)
    \]
    \[
        \texttt{AS} = 100,000 \times 1.20 \times 0.90 \times 1.05 = 113,400
    \]
    \item Determine Cloud Resource Budget, \texttt{CRB}:
    \[
    \textit{\texttt{CRB}} = \min(120,000, 113,400) = 113,400
    \]
\end{enumerate}

\begin{table}[htbp]
    \centering
    \caption{Cloud Resource Budget Calculations with and without Variability Factor}
    \begin{tabular}{|c|c|c|}
        \hline
        \textbf{\texttt{C} (in percent)} & \textbf{\texttt{CRB} (V = 0)} & \textbf{\texttt{CRB} (V = 0.1)} \\
        \hline
        0\% & $120,000$ & $120,000$ \\
        \hline
        10\% & $108,000$ & $118,800$ \\
        \hline
        20\% & $96,000$ & $105,600$ \\
        \hline
        30\% & $84,000$ & $100,800$ \\
        \hline
    \end{tabular}
    \label{tab:cloud_resource_budget}
\end{table}

This calculation indicates that the budget for cloud resources for that particular team or service for the upcoming period is \$113,400, reflecting the organization's historical spending adjusted for growth, cost control measures, and a 5\% variability factor to account for potential fluctuations in spending To further study the relationship between the Cost Control Factor \texttt{C} and the required Cloud Resource Budget \texttt{CRB}, we can plot a graph with varying values of \texttt{C} ranging of 0\% showcasing no cost savings measures to 30\% level of cost savings against the Historical Cloud Spend \texttt{HC} of \$100,000 and required budget of \$120,000 assuming no cost savings measures. 

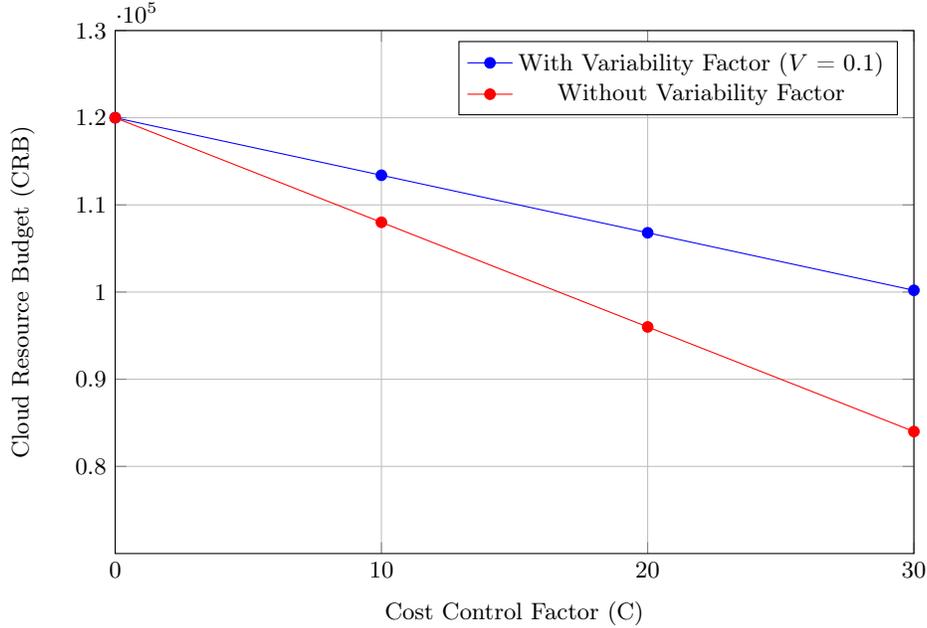
\begin{figure}[htbp]
    \centering
    \begin{tikzpicture}
        \begin{axis}[
            xlabel={Cost Control Factor (C)},
            ylabel={Cloud Resource Budget (CRB)},
            xmin=0, xmax=30,
            ymin=70000, ymax=130000,
            xtick={0, 10, 20, 30},
            ytick={80000, 90000, 100000, 110000, 120000, 130000},
            grid=major,
            width=\columnwidth, 
            height=0.7\columnwidth 
        ]
        \addplot[
            color=blue,
            mark=*
        ]
        coordinates {
            (0, 120000)
            (10, 113400)
            (20, 106800)
            (30, 100200)
        };
        \addplot[
            color=red,
            mark=*
        ]
        coordinates {
            (0, 120000)
            (10, 108000)
            (20, 96000)
            (30, 84000)
        };
        \legend{With Variability Factor ($V$ = 0.1), Without Variability Factor}
        \end{axis}
    \end{tikzpicture}
    \caption{Relationship between Cost Control Factor and Cloud Resource Budget with and without Variability Factor}
    \label{fig:crb_graph}
\end{figure}

The graph \ref{fig:crb_graph} demonstrates the inverse relationship between the Cost Control Factor (\texttt{C}) and the Cloud Resource Budget (\texttt{CRB}). As the Cost Control Factor increases, the required budget to set aside for upcoming period decreases, highlighting the impact of cost management strategies on financial planning. Additionally, with the inclusion of \texttt{V}, which accounts for potential fluctuations in spending, the budget allocation reflects a more flexible approach to financial planning ensuring that organizations are better prepared for variability in costs and historical spending.

\subsection{ABACUS Architecture}\label{sec:abacus_architecture}

\begin{figure*}[t!]
\centerline{\includegraphics[width=\linewidth,height=8cm]{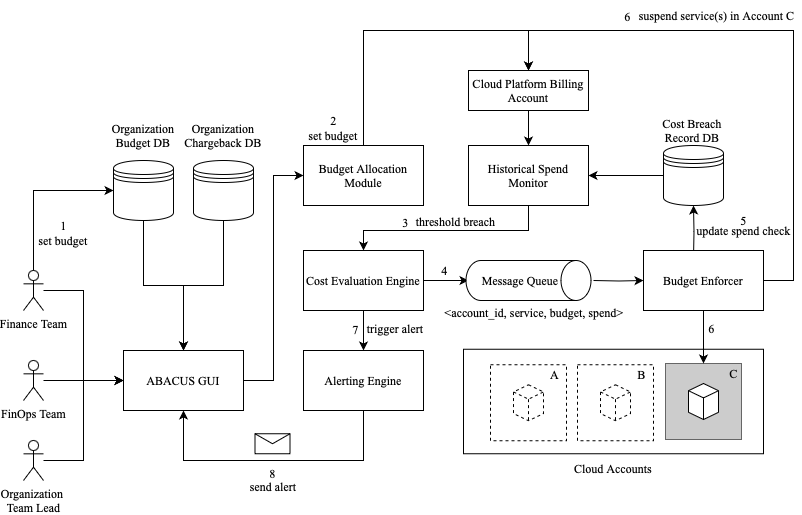}}
\caption{ABACUS Architecture}
\label{fig:arch}
\end{figure*}

The architecture showcased in figure \ref{fig:arch} provides a comprehensive foundation for managing cloud expenditures within a company. The ABACUS GUI is at the heart of this architecture, serving as the primary interface for both the Finance and FinOps teams to interact with the system. These teams are in charge of creating organizational budgets, which are then maintained in the Organization Budget and Organization Chargeback databases. The Organization Budget Database contains overall financial allocations, whereas the Organization Chargeback Database maintains cost data divided by team or project, allowing for cost attribution and accountability.

Once budget allocations have been made, the Budget Allocation Module distributes them among multiple cloud accounts maintained by the Cloud Platform Billing Account. The Cloud Platform Billing Account is a centralized component that monitors all cloud-related charges in real time. This module guarantees that each account stays within its budget, so setting a financial boundary.

The Historical Spend Monitor continuously compares expenditure data to budget restrictions. It uses a threshold-based system in which certain expenditure triggers are configured. As actual spending increases, this module continuously monitors for potential threshold breaches, allowing for proactive budget management. If the spend reaches a predetermined level, the Cost Evaluation Engine is activated to conduct a thorough examination of the expenditure patterns. This component investigates aspects such as the services that contribute to high costs, the rate of spending, and any abnormalities that may suggest inefficient resource consumption or unplanned costs.

When the Cost Evaluation Engine detects a significant threshold breach, it transmits essential data to a Message Queue, including the account ID, service type, budget, and spending information. This queue acts as a buffer, ensuring that enforcement actions are processed sequentially, avoiding conflicts and delays. This data is then collected by the Budget Enforcer, who serves as an enforcement agent. Based on the information obtained, the Budget Enforcer may halt particular cloud services or impose limits on the affected accounts. For example, in the diagram, Account C is depicted as having ceased services, indicating a budget infraction.

The Budget Enforcer consults the Cost Breach Record Database, which keeps a complete record of all spending breaches and any measures taken. This database is essential for not just rapid enforcement, but also long-term financial analysis and compliance assessments. The system provides useful data by keeping a historical record of budget breaches and the resulting enforcement measures, which may be used to identify patterns, evaluate policy efficacy, and enhance future budgeting procedures.

Simultaneously, the Alerting Engine sends real-time messages and alerts whenever a budget breach is discovered and punished. These warnings are delivered to both the Finance and FinOps teams via multiple channels, including email, Slack, and dashboard notifications. The Alerting Engine sends real-time alerts each time spending crosses the 50\%, 75\%, 90\%, 100\%, and the threshold of the budget limit. The alerting mechanism guarantees that stakeholders are swiftly notified, allowing them to take corrective action or explore the underlying causes of excessive spending.

Furthermore, the architecture enables scalability and adaptation. The ABACUS GUI can expand to accept more accounts or budget categories as the organization grows. In more specific technical terms, this expansion can be easily performed via Google Service Account IAM roles in Google Cloud Platform (GCP) or assuming IAM roles in new accounts in Amazon's AWS. The system is also intended to handle larger amounts of data, with the Message Queue guaranteeing that enforcement operations are managed efficiently even when under heavy pressure. The Cost Evaluation Engine can also be upgraded, as mentioned in the Future Work section of this paper, with machine learning algorithms to improve spending projections and provide advanced insights, such as detecting unused resources or advising budget adjustments based on consumption trends.

Finally, ABACUS can be extended by integrating advanced policy-as-code frameworks like Open Policy Agent (OPA) and HashiCorp Sentinel into the Infrastructure-as-Code (IaC) workflow to greatly improve cost management during cloud deployments. Engineering teams can reduce the risk of exceeding the budget threshold by developing policies that analyze proposed infrastructure changes in relation to financial limitations. For example, an OPA policy can be developed to assess Terraform plans and reject changes that would result in expenses beyond a predetermined level \cite{open_policy_agent_2024}. Similarly, a Sentinel policy can combine cost estimates produced by Infracost, which offers comprehensive pricing depending on Terraform configurations, to verify that total expected expenses are under budget \cite{infracost_2024}. Integrating these technologies into a CI/CD pipeline enables automated cost assessments before deployment, resulting in a strong governance mechanism.

Overall, this architecture enables a more proactive and automated approach to cloud cost management. It combines real-time monitoring, enforcement, and alerting to provide a strong solution that enables enterprises to maintain financial control over cloud expenditures, improve resource consumption, and support long-term cloud adoption initiatives. The system improves accountability and facilitates informed decision-making throughout the company by keeping detailed logs and allowing for swift reactions to budget breaches.

\section{Future Work}\label{sec:future_work}
While the existing design efficiently tackles the main difficulties of cloud cost management through real-time monitoring, budget enforcement, and alerting, there are various paths for future upgrades to enhance the system's resilience, scalability, and intelligence. This section discusses a few major future work items for ABACUS.

\subsection{Integration of Machine Learning for Predictive Budgeting}
Incorporating machine learning (ML) models into the Cost Evaluation Engine could improve the present framework and allow for proactive budget management. Currently, budget enforcement is reactive, responding only after breaches have occurred. By including predictive algorithms for forecasting, such as time-series models (e.g., ARIMA, LSTM), the system could predict when budgets are likely to be depleted based on previous and current data \cite{arima_lstm}. Anomaly detection algorithms (such as Isolation Forest and Autoencoders) could spot anomalous expenditure patterns early, allowing for prompt responses. Furthermore, regression models can evaluate the impact of various usage parameters on costs, whereas clustering models can disclose specific usage trends. These ML-driven additions would allow for budget modifications and optimizations before breaches occurred, improving the system's responsiveness and accuracy.

\subsection{Adaptive Enforcement Policies}
Future iterations of the Budget Enforcer will include adaptive enforcement policies. Instead of using static enforcement actions such as stopping services, the system could tailor enforcement actions to business-critical workloads or Service Level Agreements (SLAs) \cite{beyer2016site}. Instead of immediately suspending a crucial production service, the system could send warnings, slow non-essential services, or impose interim spending constraints to allow critical applications to continue functioning. This dynamic method would increase flexibility while ensuring that enforcement measures do not disrupt critical activities.

\subsection{Scalable Multi-agent System-Based Optimization}
Including multi-agent systems in the cost optimization framework can considerably improve efficiency, scalability, and resilience. The solution achieves dynamic, real-time cost control by utilizing intelligent agents that monitor resource utilization autonomously and enforce budget policies across several cloud accounts \cite{deochake2024cloud}. By leveraging intelligent agents that cooperatively manage and predict resource needs, organizations can significantly reduce operational expenses while maintaining performance \cite{wang_li_mohajer_2024}. These agents can collaborate to optimize resource allocation depending on current demand, ensuring that budgets are spent efficiently while keeping costs low \cite{deochake_2022_bdi} \cite{deochake_mukhopadhyay_2020}. Decentralizing budget enforcement enables these agents to independently manage monitoring and enforcement for individual services, improving scalability and fault tolerance. This structure allows the system to scale easily, adapt to changing usage patterns, and recover from individual agent failures, resulting in a more resilient and adaptable design.

\subsection{Comprehensive Audit and Compliance Reporting}
Enhancing the Cost Breach Record Database to allow thorough audit trails and compliance reports will greatly help businesses such as banking and healthcare, which have stringent budget management regulations. Future versions may have compliance dashboards and automatic reporting tools integrated directly into the ABACUS GUI to assist firms in demonstrating cost control adherence. To maintain the integrity and availability of audit data, careful evaluation of storage methods is required. Cloud-based choices such as Google BigQuery, Amazon S3, and Azure Blob Storage provide scalability and security, but location can have an impact on costs. High-cost regions may increase prices, whereas low-cost regions might assist manage budgets. Implementing data redundancy and backups across several geographical areas improves reliability \cite{deochake_channapattan_steelman_2022}, but may increase storage costs, making this strategy particularly advantageous for organizations navigating complex regulations such as GDPR, HIPAA, or PCI DSS \cite{williams_adamson_2022}.

\section{Conclusion}
\label{sec:conclusion}
In conclusion, this paper has emphasized the crucial necessity for cloud cost optimization in today's changing digital landscape, as well as the importance of implementing FinOps concepts to enhance collaboration between finance and IT teams for successful resource management. The ABACUS algorithm and architecture were established, which are intended to improve budget enforcement and resource allocation through real-time monitoring and automated interventions. Future developments, such as the integration of multi-agent systems and machine learning models, were also proposed as a way to enable proactive budget management and compliance monitoring. These developments will not only improve the system's adaptability and scalability but will also allow enterprises to negotiate complicated regulatory settings while reaping the financial benefits of cloud computing.

%
%
%
\bibliographystyle{splncs03_unsrt}
\bibliography{abacus}
\end{document}